\documentclass[preprint]{elsarticle}
\usepackage{graphicx,epsfig}
\usepackage{amsmath,amssymb}
\usepackage{lineno,hyperref}
\hypersetup{colorlinks=true}
\usepackage{graphicx}
\usepackage{wrapfig}
\usepackage{caption}
\usepackage{subfigure}
\modulolinenumbers[5]
\biboptions{sort&compress}

\newcommand{\dd}{\mathrm{d}} 
\newcommand{\ii}{\mathrm{i}} 
\newcommand{\ee}{\mathrm{e}} 
\newcommand{\eee}[1]{\mathrm{e}^{#1}} 
\newcommand{\pd}[2]{\frac{\partial #1}{\partial #2}} 
 
\renewcommand{\phi}{\varphi}

\newcommand{\sgn}{\mathrm{sgn}}

\hypersetup{
 colorlinks=true,
 linktoc=all,
 linkcolor=red,
 citecolor=blue,
 urlcolor = blue}

\journal{Physics Letters A}

\begin{document}

\begin{frontmatter}

\title{Wave diffraction by a cosmic string}

\author[fqa,iccub]{Isabel Fern\'{a}ndez-N\'{u}\~{n}ez}
\author[fqa]{Oleg Bulashenko}

\address[fqa]{Departament de F\'{i}sica Qu\`{a}ntica i Astrof\'{i}sica}
\address[iccub]{Institut de Ci\`{e}ncies del Cosmos (ICCUB) \\ Facultat de
F\'{i}sica, Universitat de Barcelona, Mart\'{i} i Franqu\`{e}s 1, E-08028
Barcelona, Spain.}

\begin{abstract}
We show that if a cosmic string exists, it may be identified through
characteristic diffraction pattern in the energy spectrum of the observed
signal. In particular, if the string is on the line of sight, the wave field is
shown to fit the Cornu spiral.
We suggest a simple procedure, based on Keller's geometrical theory of
diffraction, which allows to explain wave effects in conical spacetime of a
cosmic string in terms of interference of four characteristic rays.
Our results are supposed to be valid for scalar massless waves, including
gravitational waves, electromagnetic waves, or even sound in case of condensed
matter systems with analogous topological defects.
\end{abstract}


\end{frontmatter}


\section{Introduction}

Topological defects may appear naturally during a symmetry-breaking phase
transition in various physical systems. One of the examples is a cosmic string
-- a long-lived topologically stable structure that may have been formed at
phase transitions in the early Universe
\cite{kibble76, vilenkin-shellard94, hindmarsh95}.
Cosmic strings are analogous to other linear defects found in condensed matter
systems: vortex lines in liquid helium \cite{williams99}, flux tubes in type-II
superconductors \cite{witten85}, disclinations in liquid crystals
\cite{bowick94}, in graphene \cite{vozmediano10},
or in metamaterials \cite{mackay10-a,mackay10-b,smolyaninov-3}.

The spacetime around a straight cosmic string is locally flat, but it globally
has a conical topology that can give rise to a variety of observable phenomena
\cite{vilenkin-shellard94, hindmarsh95}.
The most evident way to detect cosmic strings is by means of gravitational
lensing. The conical topology should produce double images of a distant
source situated behind the string \cite{vilenkin81}. The images should be
undistorted but they may overlap if the split angle, which is proportional to
the string tension, is small.
In such a case, the wave effects are extremely important as a probe in
gravitational lensing \cite{deguchi86}, that was extensively studied for compact
or point-like objects \cite{schneider92}, but only a few studies are known for
the strings \cite{linet86,osipov95,yamamoto03,suyama06,yoo13}.

In this Letter we show that the wave propagation in conical spacetime,
caused by a cosmic string or similar topological defects,
can be effectively treated in the framework of the celebrated Arnold
Sommerfeld's half-plane diffraction problem
\cite{sommerfeld1896,sommerfeld54,sommerfeld04}.
In this way, we find analytical solutions in terms of Fresnel integrals, that
let us to conclude that the wave effects in conical space are determined by
a unique parameter, the Fresnel zone number.
For the wave effects to be detectable in a compact-mass gravitational lens, the wavelength $\lambda$ should be comparable with the
Schwarzchild radius $R_{\rm s}$ of the lens \cite{deguchi86}.
This condition cannot be applied to a string, a non-compact object with conical
topology. Instead, we show that the diffraction effects caused by a string are
of the leading order with respect to geometrical optics whenever the observation
point (either in space or in frequency spectrum) belongs to the low-number
Fresnel zone. This is in contrast to the case of a compact-mass lens, for which
diffraction scales like $O(\lambda/R_{\rm s})$.
Basing on Keller's geometrical theory of diffraction \cite{keller62}, we suggest
a simple procedure how the geometrical-optics approximation can be ``improved''
by adding just two additional paths corresponding to diffraction.
These are waves coming from the source to the observer but hitting the string
following the shortest path. In this way, the interference effects will be taken
into account to the leading order.
Our results imply that if a cosmic string exists, it may be identified through a
characteristic diffraction pattern in the energy spectrum of the observed
signal. Finally, we show that if the string is on the line of sight, the wave
amplitude fits the Cornu spiral -- the prominent result for the Fresnel
diffraction by a straight edge or a slit.

\section{Spacetime of a cosmic string}

\begin{figure}[h!]
\centering
\includegraphics[width=0.65\columnwidth]{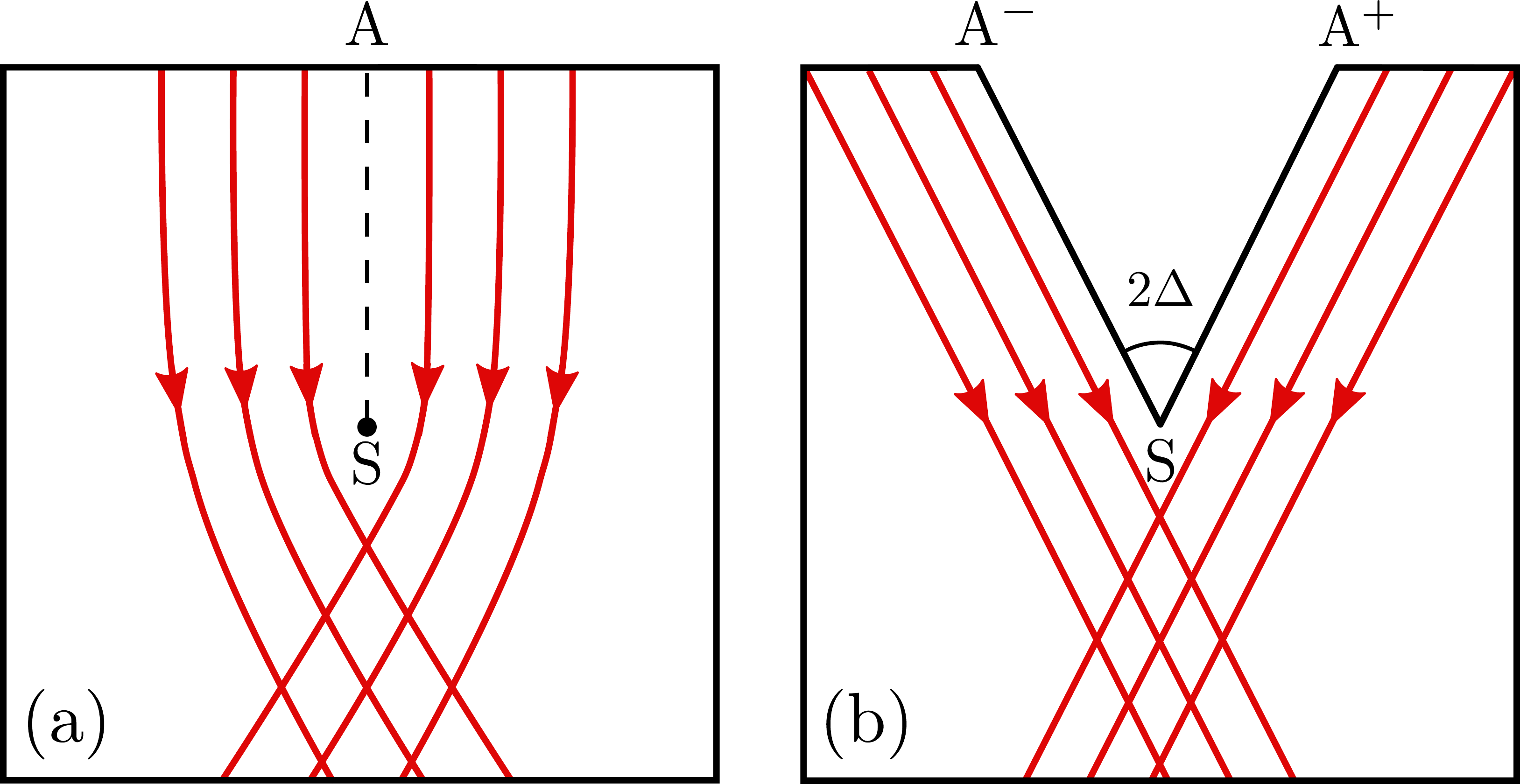}
\caption{Geodesics in the conical space on $z$=$0$ plane:
(a) curved spacetime; (b) flat spacetime with a deficit angle $2\Delta$.
The cut half-plane $SA$ is perpendicular to the plane of the figure with the
edge $S$ coinciding with the string. After the angular transformation the
half-plane $SA$ is converted to a wedge of two half-planes $SA^-$ and $SA^+$,
which should be identified.}
\label{fig:bound}
\end{figure}

We start with a spacetime metric for a static cylindrically symmetric cosmic
string \cite{vilenkin81,gott85}
\begin{equation}
\dd s^2= -\dd t^2 + \dd r^2 + (1-4G\mu)^2r^2\dd \phi^2 + \dd z^2,
\label{eq:metric-curv}
\end{equation}
where $G$ is the gravitational constant, $\mu$ is the linear mass density of the
string lying along the $z$-axis, ($t,r,\phi, z$) are cylindrical coordinates,
and the system of units in which the speed of light $c=1$ is assumed.
With a new angular coordinate $\theta=(1- 4G\mu) \phi$,
the metric \eqref{eq:metric-curv} takes a Minkowskian form
\begin{equation}
\dd s^2= -\dd t^2 + \dd r^2 + r^2\dd \theta^2 + \dd z^2,
\label{eq:metric-flat}
\end{equation}
which is locally flat, but it globally has a conical topology, since a wedge of
angular size $8\pi G\mu$ is taken out from flat space and the two faces of
the wedge are identified \cite{vilenkin81,vilenkin-shellard94}.
By introducing the deficit angle $2\Delta$ with
\begin{equation}
\Delta = 4\pi G \mu,
\label{eq:delta}
\end{equation}
the angular coordinate $\theta$ spans the range $2\pi - 2\Delta$.
Solutions of Hamilton's equations \cite{pla16} for both geometries are depicted
in Fig.\ \ref{fig:bound}.
One can see that geodesics for the metric \eqref{eq:metric-curv} are curved and
deflected an angle
$\Delta$ \cite{vilenkin81,vilenkin84}.
However, in coordinates \eqref{eq:metric-flat} they are just straight lines.
Since geodesics passing on opposite sides of the string eventually cross, one
should expect interference or diffraction effects.

\section{Wave equation in conical space}

We consider the question of finding a solution of the wave equation in
background \eqref{eq:metric-curv} corresponding to a time harmonic distant
source, so that the incident waves are plane waves.
In order to reduce the problem to two dimensions, the waves are assumed
to be emitted in the direction orthogonal to the string.
Similarly to Ref.\ \cite{suyama06}, we write the wave equation for a scalar
field
$U(r,\phi)$ as
\begin{equation}
\left(
\pd{^2}{r^2}+\frac{1}{r}\pd{}{r}+\frac{1}{\beta^2 r^2}\pd{^2}{\phi^2}
+\omega^2 \right)U=0,
\label{wave-eqn}
\end{equation}
where we denoted $\beta=1-\Delta/\pi$.
We assume that Eq.\ \eqref{wave-eqn} is valid for electromagnetic waves,
as well as for gravitational waves when the effect of gravitational lensing on
polarization is negligible and both types of waves can be described by a scalar
field
\cite{misner}.
A plane wave of unit amplitude incident from the direction $\phi_0$ is described
by
\begin{equation}
U=\ee^{\ii k r \cos\{\beta(\phi-\phi_0)\}}.
\label{inci}
\end{equation}
Next, unlike Ref.\ \cite{suyama06}, we perform the coordinate transformation
taking advantage of the flat background \eqref{eq:metric-flat}.
To do that, we place the cut line $SA$ strictly perpendicular to the wavefront
of the incident wave, as shown in Fig.\ \ref{fig:bound}(a), so we get
$\partial_\phi U=0$ at the cut.
Then, we assign the values $\phi_0^-=-\pi$ to the left and $\phi_0^+=\pi$ to
the right of the line $SA$.
After angular transformation $\theta=\beta \phi$,
the line $SA$ converts to the wedge $SA^-$, $SA^+$ given by the angles
$\pm(\pi-\Delta)$.
The incident field \eqref{inci} will now be represented by two plane waves
\begin{equation}
U=\ee^{\ii k r \cos(\theta\pm\Delta)}
\label{inci2}
\end{equation}
incoming from the directions $\pm(\pi-\Delta)$ with wavefronts perpendicular to
the faces of the wedge and propagating in a flat background
[see Fig.\ \ref{fig:bound}(b)].
As we are going to show now, this problem can be reduced to the canonical
problem of diffraction on a perfectly conducting half-plane screen solved
exactly by Sommerfeld \cite{sommerfeld1896,sommerfeld54,sommerfeld04}.

\begin{figure}[h!]
\centering
\includegraphics[width=0.35\columnwidth]{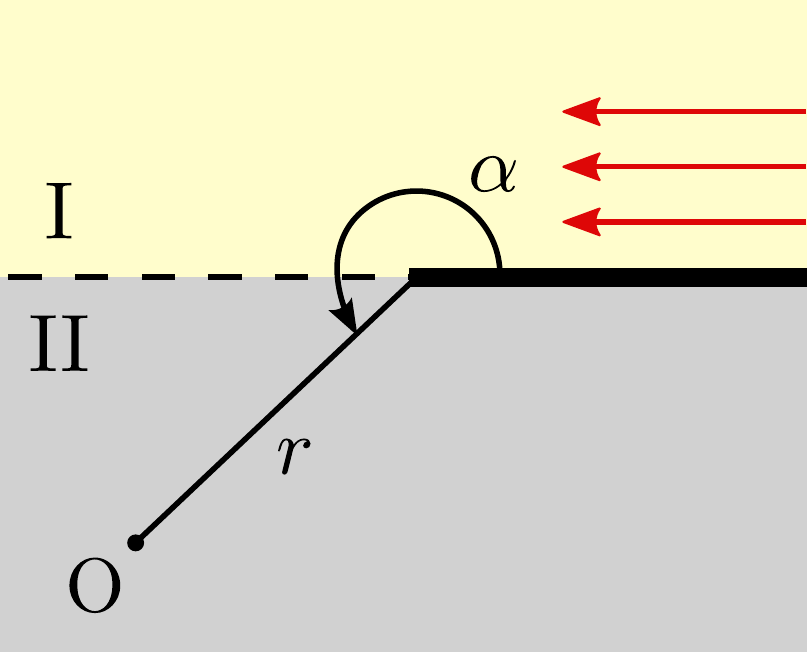}
\caption{Plane wave grazing a half-plane screen (thick line).
The entire space is split into two regions: illuminated (I), shadow (II).}
\label{fig:half-plane}\end{figure}

Let us consider a single plane wave grazing an infinite half-plane screen, as
shown in Fig.\ \ref{fig:half-plane}. Following Sommerfeld \cite{sommerfeld54},
the exact solution for the field at any point $O(r,\alpha)$ can be written in
the compact form
\begin{equation}
U=\eee{-\ii kr\cos\alpha}\mathcal{F}\left(\sqrt{2kr}\cos\frac{\alpha}{2}\right),
\label{eq:half-plane}
\end{equation}
where $r$ is the distance from the screen edge, $\alpha$ is the angle measured
from the surface of the screen facing the source, and
$\mathcal{F}(u) = e^{-\ii\pi/4}\,\pi^{-1/2}\int_{-\infty}^u \eee{\ii s^2}\dd s$
is the Fresnel integral \cite{landau-v8}. In Eq.\ \eqref{eq:half-plane} we have
taken into account the zero angle of incidence
and the Neumann boundary condition on the screen, $\partial_\alpha U (r,0)$=$0$.
It can be verified that solution \eqref{eq:half-plane} contains both the
geometrical-optics (GO) and the diffracted (D) fields.
Indeed, for the angles $0<\alpha<\pi$, in the limit $kr\to \infty$ far away from
the edge, one gets $\mathcal{F}\to 1$ and $U=\eee{-\ii kr\cos\alpha}$, which is
the GO incident field.  Whereas, for the angles $\pi<\alpha<2\pi$, one obtains
$\mathcal{F}\to 0$ giving $U=0$ at infinity.
The Fresnel function $\mathcal{F}$ smooths the discontinuity of the GO solution
across the shadow boundary $\alpha$=$\pi$ making the total field continuous
everywhere. This smooth transition constitutes the diffraction phenomenon
\cite{sommerfeld54}.
It should also be noted that the original Sommerfeld problem treats two
possible boundary conditions on the screen (Dirichlet or Neumann) depending
on the polarization of the incident field. However, for grazing incidence, only
one polarization can propagate
which corresponds to the Neumann condition. On the other hand, the zero field
condition is unphysical for the conical space we consider.

Having defined the solution for a single half-plane,
we now construct the wave field corresponding to the geometry of Fig.\
\ref{fig:bound}(b), in which we have two plane waves \eqref{inci2} grazing the
faces of the wedge.
Substituting grazing angles: $\alpha = \pi-\Delta \mp \theta$ into Eq.\
\eqref{eq:half-plane}, we obtain the total field $U(r,\theta)$ at the
observation point $O$ 
\begin{equation}
U=\eee{\ii kr\cos(\Delta+\theta)}\mathcal{F}(w^+)+\eee{\ii
kr\cos(\Delta-\theta)}\mathcal{F}(w^-)
\label{eq:u-diff}
\end{equation}
with $w^\pm=\sqrt{2kr}\sin[(\Delta\pm\theta)/2]$.
It describes the wave effects in the gravitational lensing by a cosmic string.
It is easy to verify that for $\Delta$=$0$, it reduces to the unlensed field
$U_0=\eee{\ii kr \cos\theta}$, which is a usual plane wave in Minkowskian space.

\begin{figure}[h]
\centering
\includegraphics[width=0.8\columnwidth]{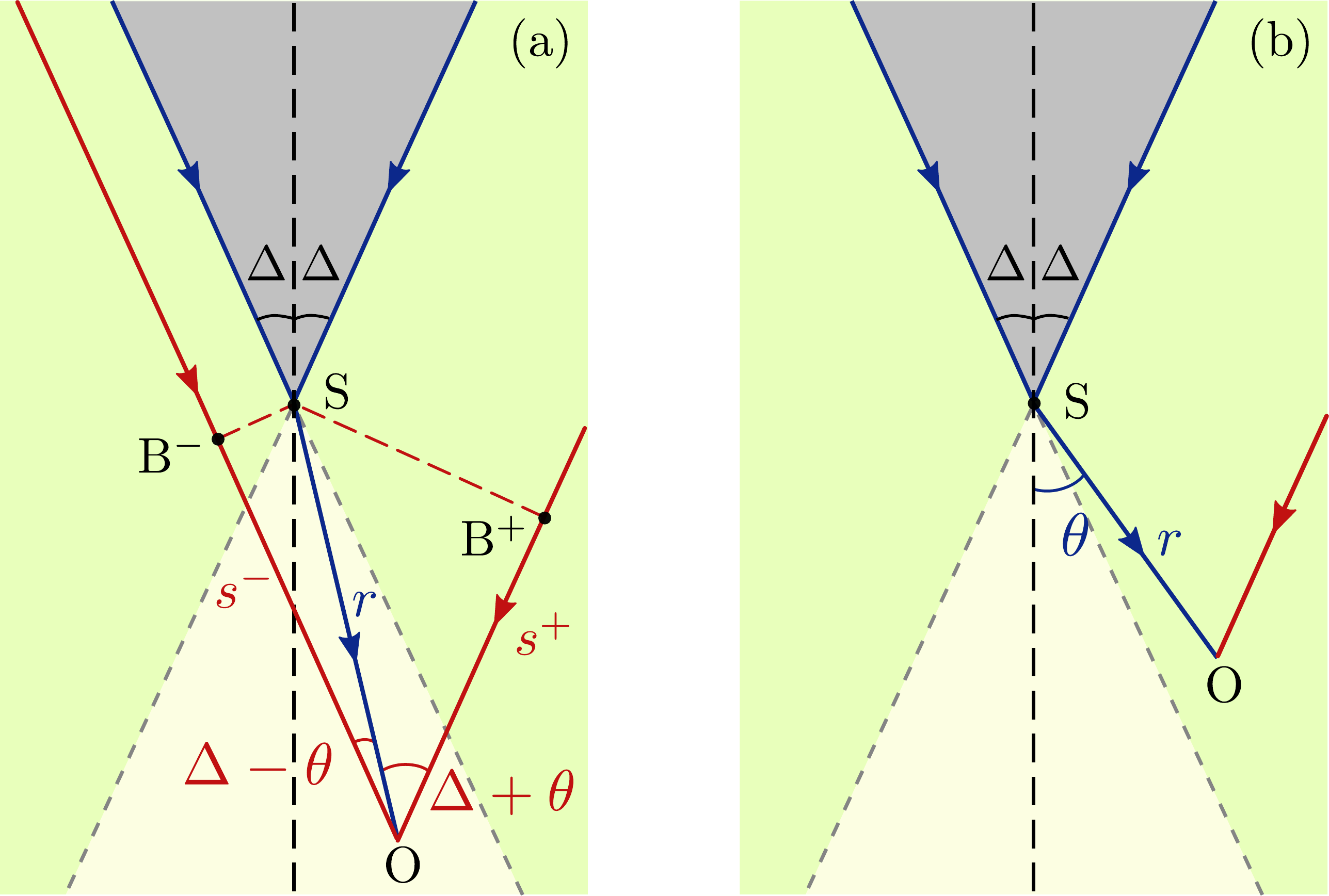}
\caption{Geometrical theory of diffraction in conical space.
The deficit wedge $2\Delta$ is shadowed.
Geometrical-optics (red) and diffraction (blue) rays determine the leading
order contribution at the observation point $O$.
The distances from $O$ to $B^\pm$ are equal to $s^\pm$.}
\label{fig:rs}
\end{figure}

For further analysis Eq.\ \eqref{eq:u-diff} can be rewritten in a more
convenient form in terms of the eikonals $s^\pm=r\cos(\Delta\pm\theta)$ of the
GO waves.
The arguments of the Fresnel function become
$w^\pm =\sigma^\pm \sqrt{k(r-s^\pm)}$ where
$\sigma^\pm \equiv \sgn(\Delta\pm\theta)$ are sign functions giving
$+1$ in the region illuminated by the corresponding GO wave and $-1$ in the
shadow.
We get
\begin{equation}
U = \eee{\ii ks^-}\mathcal{F}(\sigma^-\sqrt{\pi N^-})
+ \eee{\ii ks^+}\mathcal{F}(\sigma^+\sqrt{\pi N^+}),
\label{eq:u-rigor}
\end{equation}
with
$N^\pm=k(r-s^\pm)/\pi$, 
the number of half-wavelengths matched in the path difference $r-s^\pm$. According to geometrical optics, one can distinguish a double-imaging region
illuminated by both GO waves, $-\Delta<\theta<\Delta$,  in which two images are
seen at the observation point [see Fig.\ \ref{fig:rs}(a)]; and a single-imaging
region outside, in which one of the GO waves is shadowed
[Fig.\ \ref{fig:rs}(b)].
The obtained Eq.\ \eqref{eq:u-rigor} gives more complex spatial structure due to
diffraction. It determines Fresnel zones which consist of two families of nested
parabolas, all with a common focus at the string.
In $(r,\theta)$ polar coordinates the parabolas are given by:
$r\{1-\cos(\Delta\pm\theta)\} = (\lambda/2)\,N_F$, with their directrices
perpendicular to the lines $\theta=\mp\Delta$, respectively.
Each parabola corresponds to a constant phase difference $k(r-s^\pm)$, and the positive integers $N_F$ mark the Fresnel zones numbers \cite{fresnel-zone}.

\section{Geometrical theory of diffraction}

To proceed further, we make asymptotic expansion of the Fresnel integrals for
large arguments \cite{sommerfeld54}
\begin{equation}
\mathcal{F}(w) = \mathcal{H}(w) -
\frac{\eee{\ii\pi/4}}{2\sqrt{\pi}}\, \frac{\eee{\ii w^2}}{w}
+ O(w^{-3}).
\label{eq:expansion}
\end{equation}
Here, $\mathcal{H}(w)$ is the Heaviside step function. Substituting in Eq.\
\eqref{eq:u-rigor}, we obtain
\begin{equation}
U = h^- \,\eee{\ii ks^-} + h^+ \,\eee{\ii ks^+}
+ (D^- + D^+)\, \frac{\eee{\ii kr}}{\sqrt{2\pi kr}}. 
\label{eq:asy-rs}
\end{equation}
The first two terms with notation $h^\pm \equiv\mathcal{H}(\Delta\pm\theta)$
describe the GO waves. The step functions guarantee that the GO waves only
contribute to the respective illuminated regions.
The third term is the leading order term of the diffracted field.
It describes a cylindrical wave emanating from the string vertex and whose
amplitude depends on the direction through the ``diffraction coefficients'':
\begin{equation}
D^\pm = - \frac{\eee{\ii \pi/4}}{2}
\frac{1}{\sin{[\frac{1}{2}(\Delta\pm\theta})]}.
\label{eq:diff-D}
\end{equation}
Note that, since the diffracted wave compensates the discontinuity of GO waves,
the coefficients $D^\pm$ must go to infinity at the corresponding shadow lines
$\theta=\pm\Delta$.
The solution \eqref{eq:asy-rs} can thus be interpreted in the framework of
Keller's geometrical theory of diffraction \cite{keller62}.
For the double-imaging region, the total wave field at a point $O$ [see Fig.\
\ref{fig:rs}(a)] is determined by the sum of: two GO rays coming from the source
to the observer directly and two D rays going from the source but hitting the
edge -- the string location -- following the shortest path (Fermat's principle
for edge diffraction
\cite{keller62}).
In the single-imaging region,
only one of the GO rays but both D ones contribute [Fig.\ \ref{fig:rs}(b)].
Notice, that D-ray terms in Eq.\ \eqref{eq:asy-rs} asymptotically
$\sim O \lbrack (N_F)^{-1/2}\rbrack$ that makes them of the leading order, along
with the GO ones, at the points located in the low-number Fresnel zones.

\section{Discussion of the results}

\begin{figure}[h!]
\centering
\includegraphics[width=0.8\columnwidth]{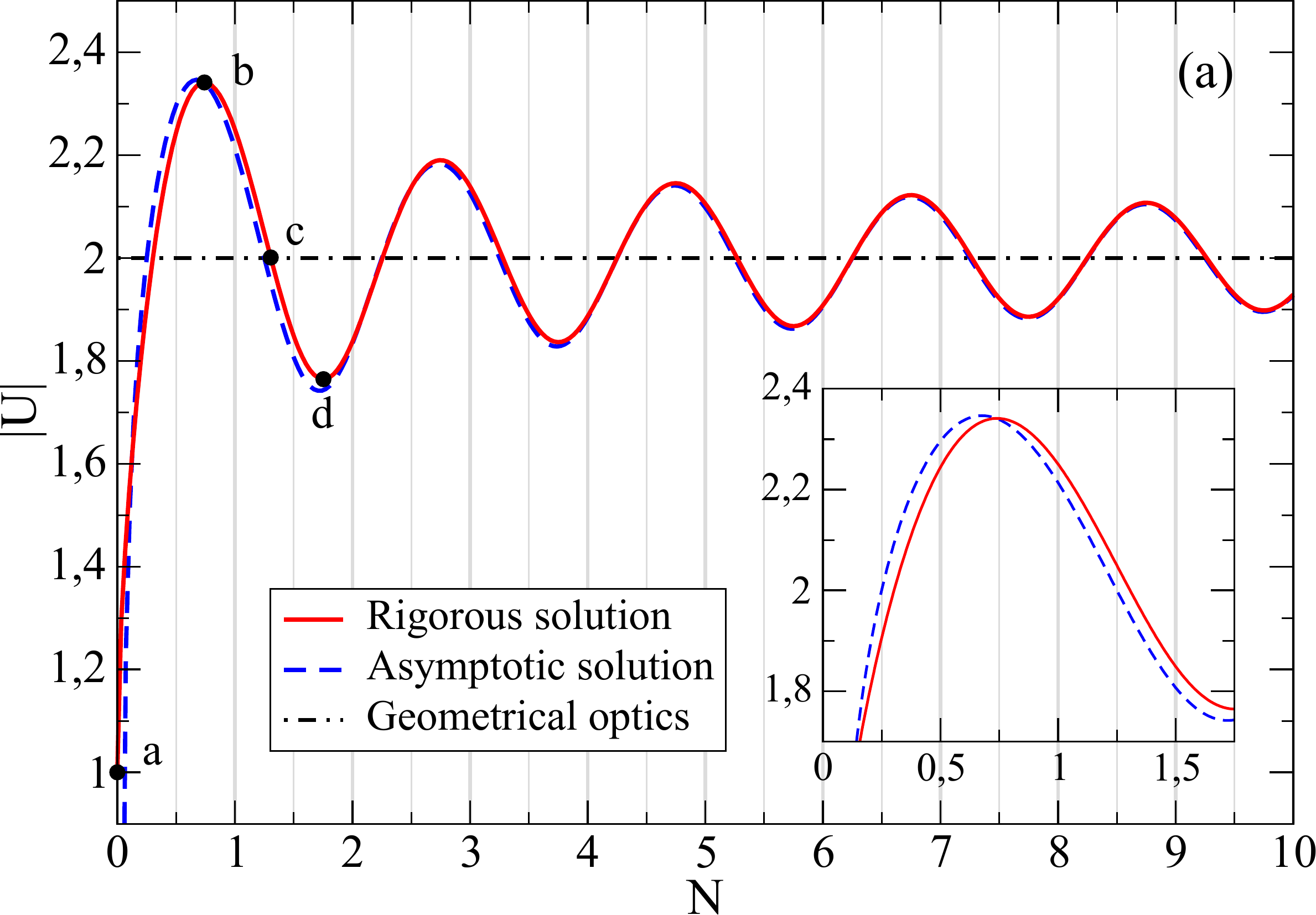}
\par\vspace*{0.25cm}
\includegraphics[width=0.8\columnwidth]{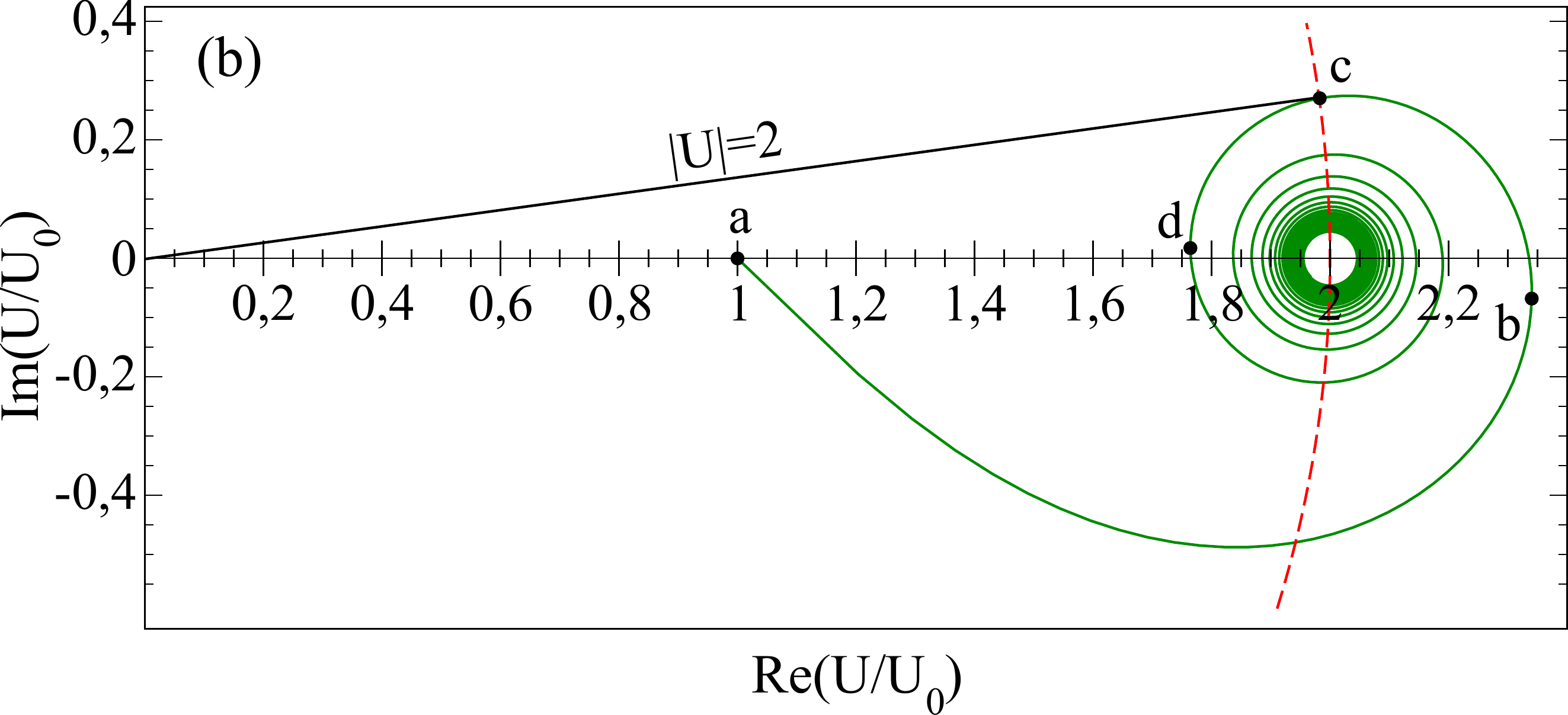}
\caption{Wave field
(a) and Cornu spiral (b) for the case when the string is on the line of sight.
The equivalent points on two figures are labeled. The cross points of circular
arc (red) with Cornu spiral correspond to $|U|$=$2$ geometrical-optics values. }
\label{fig:int-cornu}
\end{figure}

Let us analyze the obtained solution for the case when the string is on the
observer-source line of sight ($\theta=0$). Due to the symmetry, the
contributions from both sides are equal, thus the total field normalized to its
unlensed value will be of double amplitude as Eq.\ \eqref{eq:u-rigor} shows:
\begin{equation}
\left(\frac{U}{U_0} \right)_{\theta=0}  =
2\eee{-\ii \pi N} \mathcal{F}(\sqrt{\pi N}).
\label{eq:u0}
\end{equation}
We observe that finally the wave field on the line of sight is determined by
only one parameter
\begin{equation}
N = \frac{kr}{\pi}(1-\cos\Delta) = \frac{r-s}{\lambda/2},
\label{N}
\end{equation}
which is the number of half-wavelengths matched in the path difference between
the GO and D rays.
For the modulus of the field we obtain
\begin{equation}
|U|_{\theta=0}=2 | \mathcal{F}(\sqrt{\pi N}) |.
\label{inten-0}
\end{equation}
The amplification by a factor of 2 is a distinguishing feature of the
gravitational lensing by string. It is related to the appearance of double
images of identical brightness when the observer is on the line of sight or
nearby \cite{vilenkin81,gott85}.
The field \eqref{inten-0} as a function of the dimensionless parameter $N$
is plotted in Fig.\ \ref{fig:int-cornu}. We see the oscillating pattern similar
to that for the straight-edge diffraction \cite{sommerfeld54,born-wolf-03}, but
with some differences: for the diffraction by string, the shadow part of the
curve is missing and the magnitude is doubled.
The whole range of values of the field can be visualized by means of the Cornu
spiral [Fig.\ \ref{fig:int-cornu}(b)].
Here, the real and imaginary part of the wave field \eqref{eq:u0} are depicted
for different values of $N$. (We omit the phase factor $\eee{-\ii \pi N}$ which
does not contribute to the absolute value.)
In the figure, the length of the cord between the origin and any point of the
spiral is $|U|$ and the length of the Cornu curve between the end point $a$
($N$=$0$) and one on the spiral is $\sqrt{\pi N}$. So, the spiral represents
a mapping of real $N$-axis on the complex $\mathcal{F}$-plane.
As $N\to\infty$, both the wave field in Fig.\ \ref{fig:int-cornu}(a) and the
spiral approach the value of 2, the limiting GO value.
Some illustrative points in Fig.\ \ref{fig:int-cornu} have been highlighted.
To get more insight into the solution \eqref{inten-0}, one can use the
asymptotic expansion \eqref{eq:expansion}. We get
\begin{equation}
|U|_{\theta=0} \approx 2 \left[ 1-\frac{1}{\pi \sqrt{N}}
\cos{\left( \pi N + \frac{\pi}{4} \right)} \right]^{1/2}.
\label{asympt-0}
\end{equation}
The comparison of the rigorous and asymptotic solutions
in Fig.\ \ref{fig:int-cornu}(a) indicates a little difference, which for
$N\gtrsim 2$ becomes negligible.
Therefore, we can use Eq.\ \eqref{asympt-0} to determine the maxima and minima
of the diffraction:
\begin{equation}
N= - \frac{1}{4} + n \quad \text{with } \quad
\begin{matrix}
n=1,3,5\dots & \text{maxima},  \\
n=2,4,6\dots & \text{minima}.
\end{matrix}
\label{eq:N}
\end{equation}
The highest maximum occurs at $N_1\approx 3/4$ giving the field
$|U_{\rm max}| \approx 2 [1+2/(\pi\sqrt{3})]^{1/2} \approx 2.34$
(point \textit{b} in Fig.\ \ref{fig:int-cornu}).
This value corresponds to the intensity
$I_{\rm max} = |U_{\rm max} |^2 \approx 5.47$.
The diffraction pattern given by  Fig.\ \ref{fig:int-cornu} can, in principle,
be detected in two different ways: (i) if the observation frequency is fixed,
one can interpret the variation of field with $N$ as variation with distance $r$
between the string and the observer, since $N\sim r$. In this case, the first
maximum will be located at a distance
\begin{equation}
r_1 = N_1\, \frac{\lambda}{2}\, \frac{1}{1-\cos{\Delta}} \approx
N_1 \frac{\lambda}{\Delta^2},
\end{equation}
where the last approximation corresponds to $\Delta\ll 1$.
(ii) On the other hand, if $r$ is fixed, one could detect the field oscillations
caused by diffraction in the frequency spectrum due to $N\sim k\sim \omega$.
In this case, the maximum amplification will be detected at $\lambda^*\approx
r\Delta^2/N_1$. 
It should also be noted that the asymptotic solution \eqref{asympt-0}, which we
have obtained from the Fresnel integral, in the limit $\Delta\ll 1$ coincides
with the one obtained by Suyama et al.\ \cite{suyama06} by different approach.

\begin{figure}[h!]
\centering
\includegraphics[width=0.8\columnwidth]{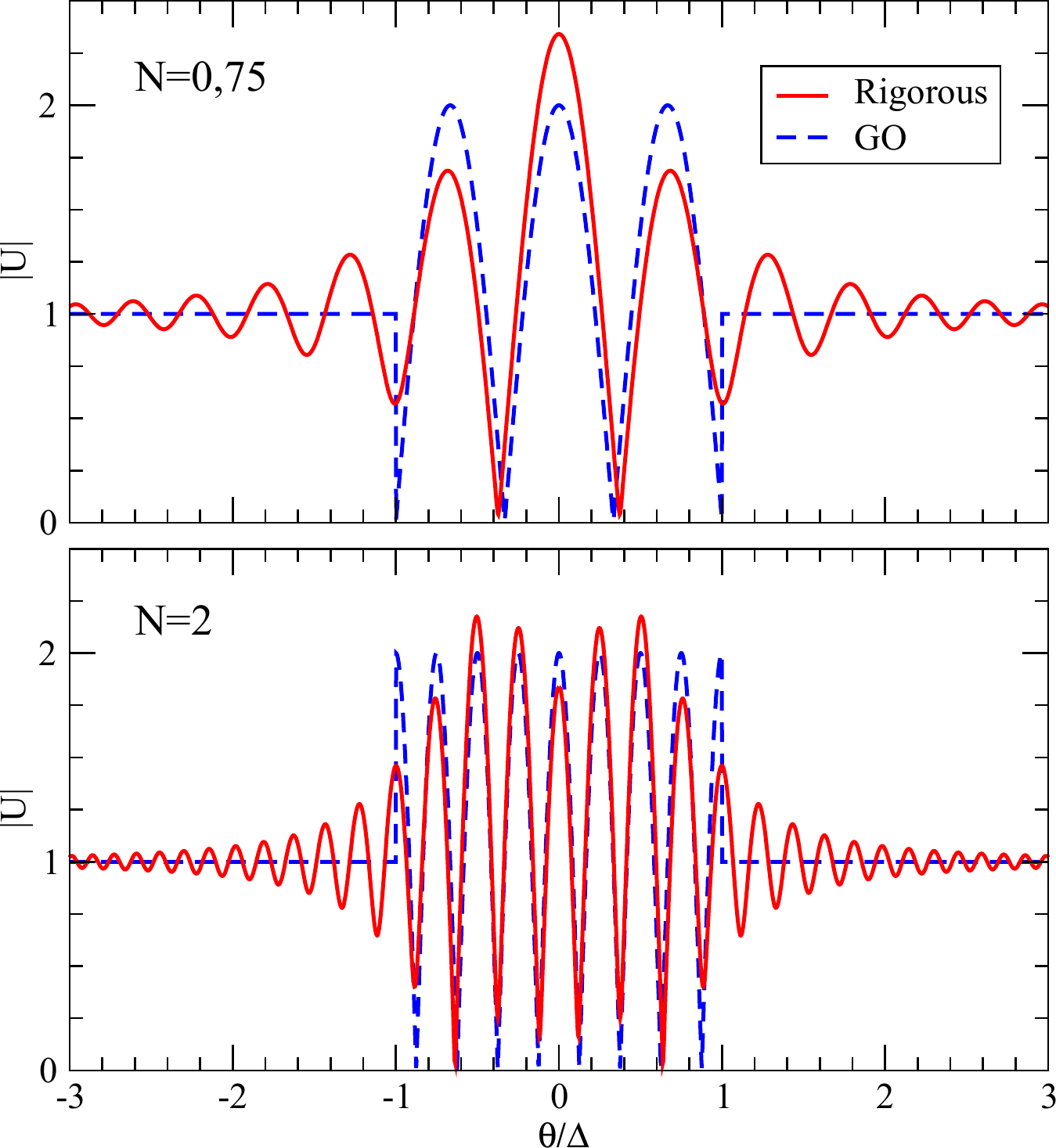}
\caption{\label{fig:int-phi}Comparison of the rigorous and geometrical optics
solution of the field $|U|$ in terms of the observation angle $\theta$
normalized to $\Delta$, for $N=0.75$ and $N=2$ with $\Delta=0.01\pi$.}
\end{figure}

When the string is not on the line of sight, $\theta\neq 0$, the wave paths from
two images are different, $s^+\neq s^-$, therefore one should expect the
interference effects (constructive or destructive) even for the GO waves (for
$\theta$=$0$ it is always constructive). If we write out the GO field from Eq.\
\eqref{eq:asy-rs} 
\begin{equation}
|U_{\rm GO}|=[h^+ + h^- +2\, h^+h^-\cos(2kr\sin\theta\sin\Delta)]^{1/2},
\label{GO}
\end{equation}
we obtain the wave oscillations amplified by a factor of 2 in the
double-imaging region, and neither interference nor amplification ($|U_{\rm
GO}|=1$) in the single-imaging region.
However, if the full solution is considered, which includes the diffraction terms,
the behavior becomes qualitatively different.
This is illustrated in Fig.\ \ref{fig:int-phi} where the total solution
\eqref{eq:u-rigor} and the GO field \eqref{GO} are shown as functions of the
angle $\theta$.
For convenience, we normalize $\theta$ by $\Delta$, so that the boundary
between the double and single-imaging regions corresponds to $\pm 1$.
We see that diffraction adds new features to the interference pattern:
(i) a higher than 2 amplification in wave magnitude;
(ii) a smooth transition in the amplitude shape when crossing the shadow
boundaries;
(iii) an amplification and oscillations in the single-imaging region.
The oscillation spacing in the central part of the pattern can be estimated
approximately from the GO field. From Eq.\ \eqref{GO} we find
\begin{equation}
 \sin{\theta_m} = \frac{1-\cos{\Delta}}{\sin{\Delta}} \frac{m}{2N}
\quad \text{with } m \in\mathbb{Z},
\end{equation}
where even $m$ determines the maxima and odd $m$, the minima.
In the limit $\Delta\ll 1$ which implies $\theta\ll 1$ for the double-imaging
region, this relation is simplified to
\begin{equation}
\frac{\theta_m}{\Delta} = \frac{m}{4N},
\end{equation}
where the $m$ values are limited by $-4N \leq m \leq 4N$.
Thus, the angular spacing between two consecutive maxima or minima will be 
$\delta\theta \approx \Delta/(2N)$.

\begin{figure}[h!]
\centering
\includegraphics[width=0.6\columnwidth]{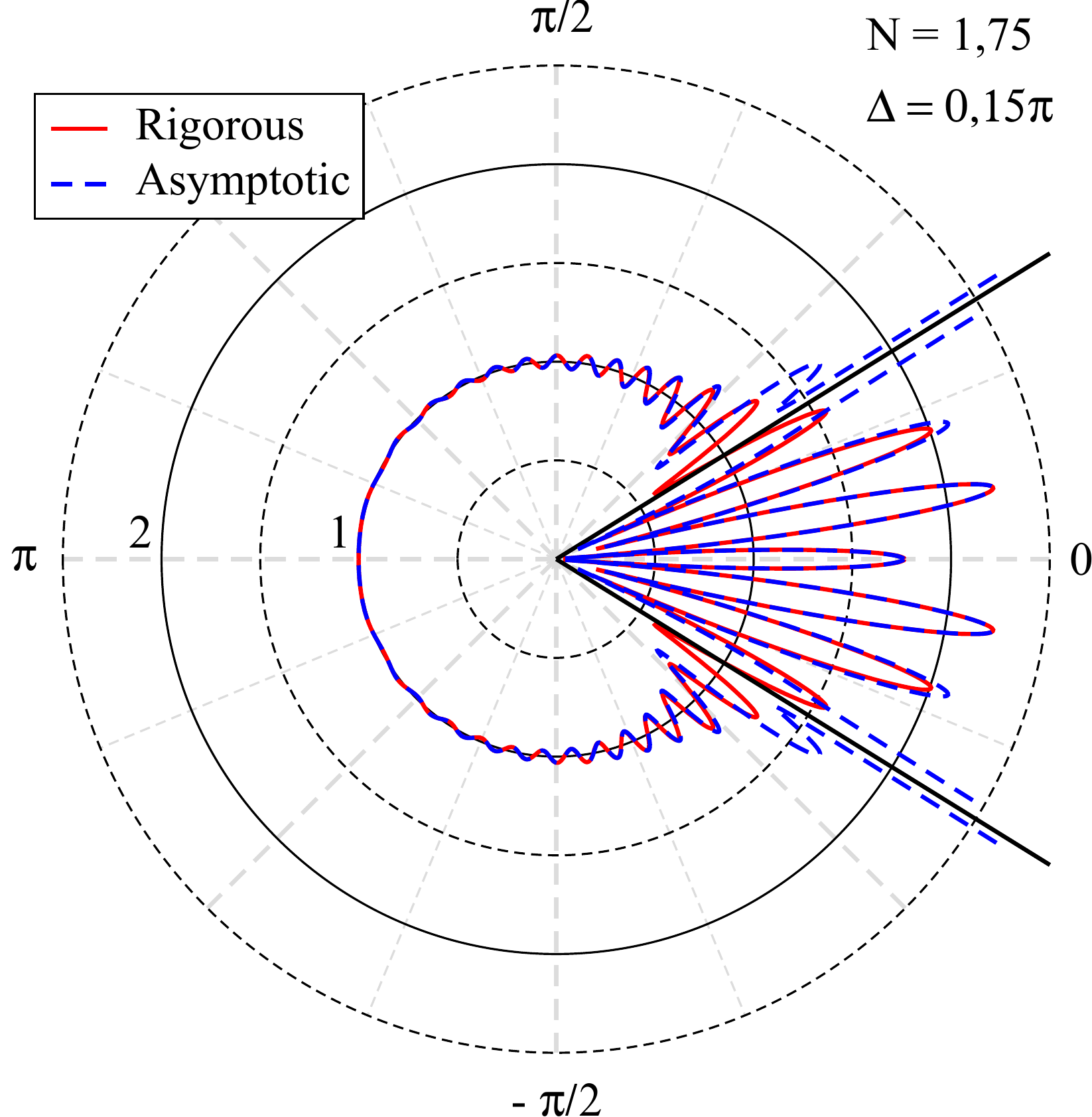}
\caption{\label{fig:polar}Angular dependence of wave field $|U|$ caused by plane
wave incident from $\phi$=$\pi$ direction in conical space
\eqref{eq:metric-curv}, as compared with asymptotic solution.
The field values correspond to fixed $kr$ determined by $N$=$1.75$ and
$\Delta$=$0.15\pi$.
The double-imaging region is bounded by two black radial lines.}
\end{figure}
Note that all the results obtained for the angle $\theta$ in flat space can be
expressed in terms of the ``physical'' angle $\phi$ in curved space by means of
the substitution $\theta=\beta \phi$.
To get more insight into the angular distribution of the field in $\phi$-space
spanning the range of $2\pi$, we present in Fig.\ \ref{fig:polar} a polar plot
for the field $|U|$ given by exact solution \eqref{eq:u-rigor} along with its
asymptotic limit \eqref{eq:asy-rs}. For a better view, a rather large value of
the deficit angle $\Delta$ is taken.
It is seen that both solutions almost coincide for all directions except at the
shadow lines $\phi=\pm \Delta/\beta$, where the diffraction coefficients
\eqref{eq:diff-D} diverge. This singularity, however, can be overcome in the
framework of so-called ``uniform theory of edge diffraction'' \cite{lee76},
which makes the asymptotic solution to be finite and continuous everywhere
including at the shadow boundaries.

\section{Conclusions and perspectives}

In conclusion, the analytical theory we have presented in this Letter, describes
propagation of scalar waves in conical spacetime created by a straight cosmic
string.
Even though the string scenario for galaxy formation requires a small deficit
angle, $\Delta\ll 1$ \cite{vilenkin-shellard94}, we believe that our results are
also applicable for condensed matter systems for which a wider range of $\Delta$
is observed.
For instance, the effective geometry \eqref{eq:metric-curv} describes
propagation of sound near a topological defect (disclination) in nematic liquid
crystals, where $c$ is the velocity of sound \cite{pereira13}.
As long as the D field decays to 0 as it approaches the boundary
$\pm\pi$ (see Fig. \ref{fig:polar}), our assumptions in the model are justified.
It would also be interesting to treat this problem as the diffraction by a cone,
or a wedge, given the geometry of the string, using different boundary
conditions and another setting \cite{dowker,fulling}.

The solutions obtained for an infinitely distant source are shown to be
determined by just one parameter, that is the Fresnel zone number.
In particular, it indicates, for a given string tension, the typical wavelength
at which one should expect the oscillations in the spectrum due to diffraction.
Taking the typical value $\Delta\sim 10^{-7}$ and a distance to the string
within our galaxy, $r\sim 10^{20}\,$m, one would obtain the highest
amplification at $\nu\approx 200\,$Hz, which is in the frequency band of Laser
Interferometer Gravitational-wave Observatory (LIGO).
The derivation can proceed in a similar way when the source is located at a
finite distance. In this case, the parameter $N$ will include dependence on the
source-string distance and the Fresnel-zone structure will be determined by
nested hyperbolas \cite{fresnel-zone}.

Basing on geometrical theory of diffraction, we suggest an intrinsically simple 
method that allows to explain diffraction phenomena in spacetime generated
by a cosmic string or similar topological defect within the principles of
geometrical optics. 
Namely, the wave field at an observation point is determined to the leading
order by interference of a few characteristic rays: the geometric-optics and the
diffracted ones.
This method, we believe, may be easily applied to other geometries.

\section*{Acknowledgements}

IFN acknowledges  financial  support  from  Universitat  de  Barcelona  under
the APIF scholarship.


\section*{References}

\begin{thebibliography}{99}

\bibitem{kibble76}
T. W. B. Kibble,
J. Phys. A  {\bf 9},  1387 (1976).

\bibitem{vilenkin-shellard94}
A. Vilenkin and E. P. S. Shellard, \textsl{Cosmic Strings and Other Topological
Defects} (Cambridge University Press, Cambridge, 1994).

\bibitem{hindmarsh95}
M. B. Hindmarsh and T. W. B. Kibble,
Rep. Prog. Phys. {\bf 58}, 477 (1995).

\bibitem{williams99}
G. A. Williams,
Phys. Rev. Lett. {\bf 82}, 1201 (1999).

\bibitem{witten85}
E. Witten,
Phys. Lett. B {\bf 153}, 243 (1985).

\bibitem{bowick94}
M. J. Bowick, L. Chandar, E. A. Schiff,  and A. M. Srivastava,
Science  {\bf 263}, 943 (1994).

\bibitem{vozmediano10}
M. A. H. Vozmediano, M. I. Katsnelson, and F. Guinea,
Physics Reports, {\bf 496}, 109 (2010).

\bibitem{mackay10-a}
T. G. Mackay and A. Lakhtakia,
Phys. Lett. A {\bf 374}, 2305 (2010).

\bibitem{mackay10-b}
T.H. Anderson, T. G. Mackay, and A. Lakhtakia,
Phys. Lett. A {\bf 374}, 4637 (2010).

\bibitem{smolyaninov-3}
I. I. Smolyaninov, V. N. Smolyaninova, and A. I. Smolyaninov,
Phil. Trans. R. Soc. A  {\bf 373}, 20140360 (2015).

\bibitem{vilenkin81}
A. Vilenkin,
Phys. Rev. D {\bf 23}, 852 (1981).

\bibitem{deguchi86}
S. Deguchi and W. D. Watson,
Phys. Rev. D {\bf 34}, 1708 (1986).

\bibitem{schneider92}
P. Schneider, J. Ehlers, and E. Falco,
 \textsl{Gravitational Lenses} (Springer, New York, 1992).


\bibitem{linet86}
B. Linet, Ann. Inst. Henri Poincare A {\bf 45}, 249 (1986).

\bibitem{osipov95}
D. L. Osipov, JETP Lett. {\bf 62}, 765 (1995).

\bibitem{yamamoto03}
K. Yamamoto and K. Tsunoda,
Phys. Rev. D {\bf 68}, 041302 (2003).

\bibitem{suyama06}
T. Suyama, T. Tanaka, and R. Takahashi,
Phys. Rev. D {\bf 73}, 024026 (2006).

\bibitem{yoo13}
C.-M. Yoo, R. Saito, Y. Sendouda, K. Takahashi, and D. Yamauchi,
Prog. Theor. Exp. Phys. 013E01 (2013).

\bibitem{sommerfeld1896}
A. Sommerfeld, Mathem. Ann. {\bf 47}, 317 (1896).

\bibitem{sommerfeld54}
A. Sommerfeld, \textsl{Lectures on Theoretical Physics, Vol. IV, Optics}
(Academic Press, New York, 1954).

\bibitem{sommerfeld04}
A. Sommerfeld, \textsl{Mathematical Theory of Diffraction}
(Birkh\"{a}user, Boston, 2004).

\bibitem{keller62}
J. B. Keller,
J. Opt. Soc. Am. {\bf 52}, 116 (1962).

\bibitem{gott85}
J. R. Gott, Astrophys. J. {\bf 288}, 422 (1985).

\bibitem{pla16}
I. Fern\'{a}ndez-N\'{u}\~{n}ez and O. Bulashenko,
Phys. Lett. A {\bf 380}, 1 (2016).

\bibitem{vilenkin84}
A. Vilenkin, Astrophys. J. {\bf 282} L51 (1984).

\bibitem{misner}
C. W. Misner and K. S. Thorne, and J. A. Wheeler,  \textsl{Gravitation}
(W. H. Freeman and Company, San Francisco, 1973).

\bibitem{landau-v8}
L. D. Landau and E. M. Lifshitz, \textsl{Electrodynamics of Continuous Media}
2nd ed., (Pergamon, New York, 1984).

\bibitem{fresnel-zone}
I. Fern\'{a}ndez-N\'{u}\~{n}ez and O. Bulashenko, (unpublished results).

\bibitem{born-wolf-03}
M. Born and E. Wolf, \textsl{Principles of Optics}, 7th ed. 
(Cambridge University Press, Cambridge, 1999).

\bibitem{lee76}
S. W. Lee and G. A. Deschamps,
IEEE Trans. Antennas Propag. {\bf 24}, 25 (1976).

\bibitem{pereira13}
E. Pereira, S. Fumeron, and F. Moraes,
Phys. Rev. E {\bf 87}, 022506 (2013).

\bibitem{dowker}
J. S. Dowker,
J. Phys. A {\bf 10}, 115 (1977).

\bibitem{fulling}
S. A. Fulling, C. S. Trendafilova, P. N. Truong, and J. Wagner,
J. Phys. A {\bf 45}, 374018 (2012).



\end{thebibliography}
\end{document}